\documentclass[%
 a4paper, twocolumn,11pt
amsmath,amssymb
aps,floatfix,superscriptaddress,
]
{revtex4-2}
\pdfoutput = 1

\usepackage[utf8]{inputenc}
\usepackage[english]{babel}
\usepackage[T1]{fontenc}
\usepackage{xkeyval}
\usepackage{graphicx}%
\usepackage[utf8]{inputenc}
\usepackage{tikz}
\usepackage{tikz-cd}
\usepackage{pgfplots}
\pgfplotsset{compat=newest}
\usepackage{lmodern}
\usepackage[export]{adjustbox}%
\usepackage{soul}
\usepackage{dcolumn}%
\usepackage{bm}
\usepackage{epstopdf}
 \usepackage[usenames,dvipsnames]{pstricks}
 \usepackage{epsfig}
 \usepackage{pst-grad} %
 \usepackage{pst-plot} %
 \usepackage[space]{grffile} %
 \usepackage{etoolbox} %
 \makeatletter %
 \patchcmd\Gread@eps{\@inputcheck#1 }{\@inputcheck"#1"\relax}{}{}
 \makeatother
\usepackage{booktabs}
\usepackage{boxhandler}
\usepackage{amsmath}
\usepackage{subcaption}
\usepackage{braket}
\usepackage{xcolor}
\usepackage{float}
\usepackage[normalem]{ulem}
\makeatletter
\let\newfloat\newfloat@ltx
\makeatother
\usepackage{algpseudocode}
\usepackage{algorithm}
\usepackage{amsfonts}
\usepackage{hyperref}
\hypersetup{
  colorlinks   = true,    %
  urlcolor     = blue,    %
  linkcolor    = blue,    %
  citecolor    = blue      %
}

\usepackage{multirow}
\usepackage{enumitem}

\begin{document}

\title{Decoding with alternating weighted graph sparsification for quantum low-density parity-check codes}

\author{Josu {Etxezarreta Martinez}}
\email{jetxezarreta@unav.es}
\affiliation{Department of Basic Sciences, Tecnun - University of Navarra, 20018 San Sebastian, Spain.}
\author{Francisco Garcia-Herrero}
\affiliation{Department of Computer Architecture and Automatics, Complutense University of Madrid, Madrid, Spain.}
\author{Ruben M. Otxoa}
\affiliation{Hitachi Cambridge Laboratory, J. J. Thomson Avenue, Cambridge, CB3 0HE, United Kingdom.}
\author{Reza Dastbasteh}
\affiliation{Department of Basic Sciences, Tecnun - University of Navarra, 20018 San Sebastian, Spain.}

\begin{abstract}
Real-time decoding is a fundamental necessity for building fault-tolerant quantum computers. In this work, we introduce decoding with alternating weighted graph sparsification (DAWG), an ensemble based approach that combines Tanner graph sparsification with ensembling belief-propagation (BP) decoders. The DAWG decoder runs BP decoders over the full detector error model Tanner graph for certain iterations to then transfer the posteriors to a sparsified graph in which additional rounds of BP are run. Each of the decoders in the ensemble uses a different transfer matrix to map the posterior information to the sparsified graph. We use the serial schedule for the minimum-sum BP decoders and, thus, employ Vibe decoding ideas by randomizing the schedules of each of them. The performance of the DAWG decoder is then numerically studied for bivariate bicycle codes and superdense color code circuits to be compared with other state-of-the-art decoding approaches. Importantly, the DAWG decoder shows a similar performance to Relay belief-propagation and Vibe decoding when combined with an order-$0$ Localized Statistics Decoding (LSD-$0$) post-processor. Crucially, the post-processor is called an order of magnitude less times than for Vibe with the same ensemble size, indicating that the introduced latency tail will be minimal. Furthermore, LSD-$0$ operates on a sparsified decoding graph of a fraction of the size of the full detector error model, which simplifies the resources required by specialized hardware to support it. Overall, the DAWG+LSD-$0$ decoder represents a viable low-cost BP-based decoding algorithm for quantum error correction codes.

\end{abstract}

\maketitle
\section{Introduction}
The construction of a fault-tolerant quantum computer is the Holy Grail of the field \cite{megaquop}. In this regard, substantial experimental progress has been made with the first demonstration of a surface-code logical qubit below the surface code threshold on a superconducting processor by Google Quantum AI \cite{googlesurface} or the integration of magic state distillation in a neutral atom processor by QuEra \cite{msdQuera}. However, significant challenges need to be overcome for the practical realization of fault-tolerance to become a reality. Real-time decoding resides as one of the most important tasks needed to be solved within the stack required to operate a fault-tolerant quantum computer at scale \cite{Battistel_2023,googlesurface}.

In this context, the design and integration of decoding algorithms is one of the most demanding aspects to reach full fault tolerance. Informally, a decoder is a classical algorithm in charge of estimating the most likely logical operator that the noise has introduced based on measured syndromes or detectors \cite{decoders}. These elements require stringent conditions for building error-corrected quantum computers that allow the execution of universal quantum algorithms. First of all, the algorithms should be accurate since it will define the required qubit overhead required to reach certain fault-tolerant level and should be flexible enough to be able to decode not only quantum memories, but also fault-tolerant logical operations. Furthermore, these should be of very low latency as even if the backlog problem can be dealt with parallel window techniques \cite{Skoric2023}, the latency required to decode a single window will impact in the logical clock rate of the system, which is known as reaction time \cite{mitten}. Due to the aforementioned latency constraints, these algorithms must be compact to be implemented in specialized hardware such as FPGA or ASIC \cite{hoefler2026quantumcomputersdifferentblueprint}. Finally, decoders should be able to process an extremely high throughput of streaming syndrome data. The combination of all these stringent requirements makes the craft of decoder design one of the most challenging components of the quantum computing stack.

In this context, quantum low-density parity-check (qLDPC) codes have emerged as low-overhead candidates for fault tolerance \cite{BBcodes,xu2023constantoverheadfaulttolerantquantumcomputation,scruby2024highthresholdlowoverheadsingleshotdecodable,mitten,kasaicodes,qgpu,directionaltiles,GBcodes}. Importantly, there has been significant progress in making these proposals practical for fault-tolerant logic \cite{extractors,magicsurgery,Williamson2026,pinnacle,oratomic,tourdegross,gongentangling,cowtanSurg,shyps}. As a result of the potential that qLDPC-based schemes present, the community has made a significant effort to propose new decoding algorithms that can be suitable for those \cite{AmbiguityClusteringOSD,Joschka_decoding_across,joschkaLSD,closedbranch,DTD,tesseract,Maan2025,bpgdg,otf,relayBP,vibe,koutsioumpas2025automorphismensembledecodingquantum,mitten,colorNvidia,cascade,gari}. The belief propagation (BP) decoder, a message passing algorithm adapted from classical communication theory, is central to many of those proposals \cite{decoders,Joschka_decoding_across}. BP is a great candidate to decode qLDPC codes, but the presence of many short loops and degeneracy results in trapping sets that confuse the message passing algorithm and do not allow for convergence \cite{trapping,degen}. This is why standalone BP cannot be used for qLDPC decoding and post-processors \cite{Joschka_decoding_across,joschkaLSD,otf,vibe,koutsioumpas2025automorphismensembledecodingquantum} or modifications of the message passing algorithm \cite{relayBP,gari} are required to reach the desired logical error rates.

In this work, we introduce decoding with alternating weighted graph sparsification (DAWG), a BP ensemble based decoder for general qLDPC codes. DAWG is constructed based on the detector error model (DEM) graph sparsification technique introduced in \cite{otf} with the aim of improving accuracy and speed of BP based decoders. In its core, DAWG is built by finding many different transfer matrices (which have different weights) that relate a full DEM with a sparsified variant of it. Thus, the transferred posterior log-likelihood ratios (LLRs) to the sparsified model are different for each element in the ensemble, providing valuable diversity that breaks trapping sets upon non-convergence of BP over the full DEM. We also adopt serial BP schedule randomization ideas from Vibe decoding \cite{vibe} to provide the decoder with extra diversity. Furthermore, since the ensemble is built with two-stage BP+BP decoders, we can include extra diversity by randomizing the schedules of the BP decoder over the full DEMs and the sparsified versions. DAWG then runs over this ensemble and selects the most likely recovery provided in the ensemble based on the detector error model priors. Finally, we use an order-$0$ localized statistics decoder (LSD-$0$) post-processor to decode the shots for which DAWG has not converged, combining the posterior LLRs obtained on the sparsified model and attempting the last step over such smaller (and sparser) model.

To test the accuracy of DAWG+LSD-$0$, we decode bivariate bicycle (BB) codes and superdense color code circuits under standard depolarizing circuit-level noise. For BB codes, we show that DAWG is able to reach logical error rates that are more than an order of magnitude lower than the ones achieved by BP+LSD-$0$ when decoded over uncorrelated DEMs. Crucially, DAWG+LSD-$0$ reaches the same performance as RelayBP for the Gross code with parameters $[[144,12,12]]$ without requiring extensive parameter optimization. For superdense color codes, DAWG+LSD-$0$ matches the performance shown by the Vibe decoder, the leading color code decoder at the moment \cite{vibe}, while requiring a significantly smaller ensemble to do so. Note also that since the post-processor runs over a significantly smaller and sparser decoding graph, the hardware resource to implement it will also be way more conservative. This makes DAWG a more compact approach for specialized hardware integration. We also analyze the convergence of DAWG when compared to Vibe decoding for both BB and superdense color code circuits and determine that the amounts in which the LSD-$0$ post-processor must be invoked to clean up unconverged syndromes is a fraction smaller. Thus, the actual latency tail introduced by DAWG coming from the post-processor will be significantly smaller, which results in a decoder with smaller latency. To provide a preliminary estimate of the benefits of DAWG for real time decoding, we discuss how it can be integrated in FPGA platforms. We report estimated latencies and observe that the method provides substantial latency improvements, specially for the post-processor that runs over the sparsified DEM. Overall, the DAWG+LSD-$0$ decoder represents a viable BP ensemble-based approach to decode general qLDPC codes.

\section{Detector error model sparsification and BP+BP decoders}
The DEM sparsification technique for general qLDPC codes was introduced in \cite{otf} as a way to improve the decoding performance of the ordered Tanner Forest (OTF) post-processor that relies on the decimation of DEMs to acyclic graphs. Since the depth of the decimated trees depends on the sparsity of the DEM, by sparsifying it to the DEM of an almost phenomenological model helps OTF to solve errors of higher weight. However, the authors observed that combining BP runs over the full DEM to then transfer the posterior LLRs to the sparsified model by means of a transfer matrix to then run some more iterations over the sparsified model significantly helps in BP convergence and success. Thus, we build upon this type of two-stage BP+BP algorithm to construct DAWG.

DEM sparsification routines can be understood as a generalization of how matching decoders deal with decoding surface codes at the circuit-level. Circuit-level DEMs of surface codes present correlated errors that do not introduce pair like detectors in the decoding graph \cite{beliefmatching}. Therefore, matching decoders decompose those correlated errors into matching-like errors in the DEM by combining the priors of those errors into the ones that compose them. In the case of belief-matching, the posterior LLRs of the BP run are then mapped to said matching DEM by means of a transfer matrix that relates the correlated errors with the ones in which they are decomposed. These LLRs are then used to guide the matching problem. In this regard, the matching graph can be understood as a sparsified version of the full correlated DEM.

The DEM sparsification technique then applies the following logic for qLDPC decoding. The main idea refers to mapping the full DEM of the code at the circuit-level to a sparsified graph that represents the same code. In \cite{otf}, the sparsified DEM is the model obtained for a circuit-level noise model that considers the entangling gates to be perfect, i.e. a quasi-phenomenological error model \footnote{Note that in a phenomenological noise model the check qubits are considered error free.}. However, in principle, one can select any matrix that shares the column space as the original DEM, i.e. $\text{IMAGE(}\mathrm{H_{dem}})=\text{IMAGE(}\mathrm{H_{sdem}})$. It is also important to have an appropriate representation of the logical observables associated with the sparsified DEM to know which logical class the decoded error refers to.

Once a sparsified model is selected, the way in which the errors in the full DEM relate to the ones in the sparsified one must be found. This is done by means of a so called transfer matrix, $A_{tr}$, that relates both models as $\mathrm{H_{dem}} = \mathrm{H_{sdem}} \cdot A_{tr}$, where the product is assumed to be modulo 2. Note that this matrix similarly captures how to relate both logical observable matrices. The way in which the transfer matrix was found in \cite{otf} was by means of an exhaustive search over a restricted radio selected by means of columns that share non-trivial positions.

\begin{figure*}[ht!]
    \centering
    \includegraphics[width=0.7\linewidth]{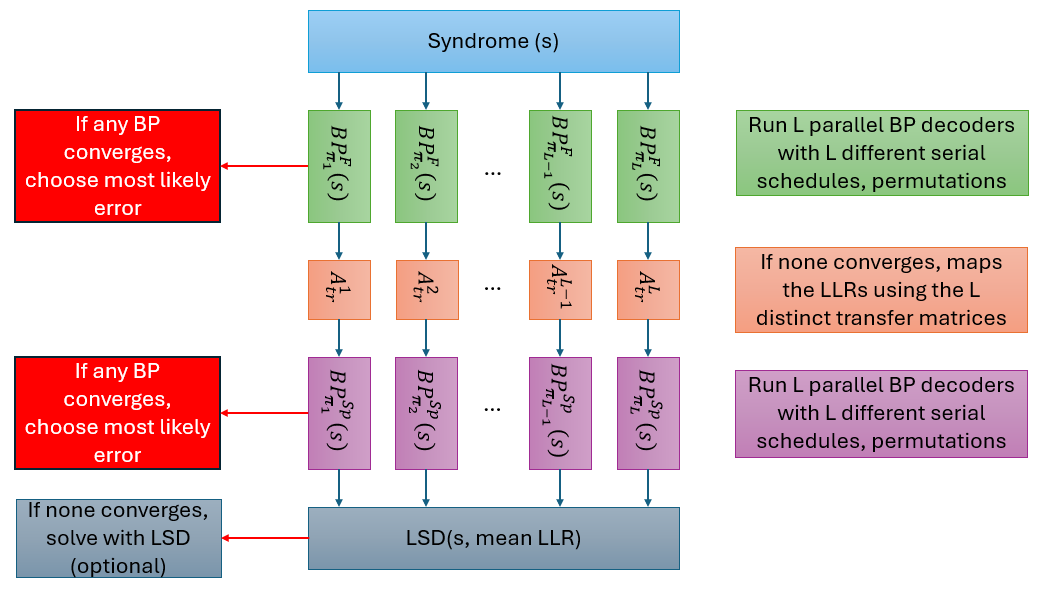}
    \caption{The DAWG+LSD-$0$ decoder. The decoder operates three stages: a) Ensemble of randomized serial BP runs over the full DEM, b) Ensemble of randomized serial BP runs over the sparsified DEM with propagated LLRs using different transfer matrices and c) LSD-$0$ with normalized sum of posterior LLRs from the sparsified DEM runs. Each of the stages is called if the previous ones did not converge. Upon convergence, the most likely error across the ensemble is chosen as a solution based on the prior probabilties.}
    \label{fig:dawg}
\end{figure*}

As said before, this type of transfer matrix is what matching decoders usually employ in order to map the priors (or BP LLR posteriors) to the matching graph. In belief-matching, the authors sum the posterior probabilities (obtained from the LLRs) by means of their transfer matrix \cite{beliefmatching}. However, an update rule to the sparsified variable nodes must take into account that pairs of errors combined into the same node cancel modulo 2. Thus, the authors in \cite{otf} defined a more refined update rule based on the XOR as

\begin{align}\label{eq:mapLLR}\nonumber
    p(e_{sdem}^i) &= 
    p\left(\bigoplus_{k\in \{j : A_{tr}^{ij}\neq 0 \}} e_{dem}^k\right) \\ & =  \frac{1}{2}\left(1-\prod_{k\in \{j : A_{tr}^{ij}\neq 0 \}} (1-2p(e_{dem}^k))\right),
\end{align}
where $p_{sdem}$ and $p_{dem}$ refer to the probabilities of the sparsified and full DEM respectively, while $A_{tr}$ is the transfer matrix. It is noteworthy to state that the previous update rule from the DEM variable nodes to the sparsified DEM one is conceptually similar to a check node update in sum-product belief propagation, i.e. the probability of the XOR of the node values \cite{decoders,Joschka_decoding_across}.

Note how one could, in principle, update the priors of the sparsified model from the DEM ones to directly decode over the smaller model. However, this leads to a significant performance loss as it was discussed in \cite{otf} considering BB codes decoded with BP plus ordered statistics decoding (OSD). Note how this is consistent with standard matching showing higher logical error rates than belief-matching \cite{beliefmatching}. Therefore, it is more convenient to run certain BP iterations over the full DEM and pass the posterior LLRs using the update rule discussed upon unconvergence. This makes the sparsified noise model to get valuable information from the BP runs over the full DEM, making it aware of the fact that the actual noise model is not exactly the one that is being decoded. This is exactly what was observed, since the BP+BP+OSD-$0$ decoder matched the logical error rate as the standard BP+OSD-$0$ decoder over the full DEM, while showing three times lower latency. This indicates that considering some BP runs over the full DEM is necessary to obtain good decoding performance.

It is important to comment that the latency improvement shown by the BP+BP+OSD-$0$ variant came from the fact that the post-processor was invoked much less times than for the original one. The main reason for it stems from the fact that combining the posterior LLRs into the sparse version that contains significantly less variable nodes breaks some trapping sets. This is because this combination of probabilities is able to break some of the equivalent solutions in the original DEM that traps BP making it to oscillate. Perturbing the LLRs to scape trapping sets has been considered in the literature \cite{degen}, but it is done in an structured manner in this proposal rather than in a random way. The DAWG decoder presented here leverages this sparsification and two stage BP ideas with ensembling to make a fast and accurate decoding proposal.

\section{Decoding with alternating weighted graph sparsification}

The transfer matrix relating the DEM with the selected sparsified version is not unique. Additionally, it is well known in the BP community that adding diversity to these algorithms results in significant decoding improvements \cite{vibe,otf,diversity}. As a result, DAWG builds upon such an idea by looking for distinct transfer matrices connecting the DEM with its sparsified version, introducing diversity in the LLRs that are mapped to the smaller decoding graphs. 

A graphical description of the full DAWG+LSD-$0$ decoder is shown in figure \ref{fig:dawg}. As observed there, the DAWG decoder proceeds as follows upon receiving a measured syndrome/detector:
\begin{enumerate}
    \item An ensemble of serial BP runs is performed over the full DEM with randomized schedules, similar to Vibe decoding \cite{vibe}. If some of the BP run converge, the most likely error is chosen among them based on the priors.
    \item If none of them converge, the posterior LLRs are mapped to the sparsified DEM selected. Each transfer matrix is selected to be different across the ensemble, ensuring a more diverse LLR propagation.
    \item Another ensembling of serial BP is run with randomized schedules over the sparsified DEMs. Similar to the first BP run, the most likely of the ensemble is selected upon convergence, based on the propagated priors to the sparsified DEMs.
    \item If no BP run converged after the DAWG stage, an LSD-$0$ post-processor over the sparsified DEM is invoked (optionally) to clean up the remaining syndromes. The input LLRs to the LSD-$0$ decoder are the normalized sum of all LLR vectors over the sparsified DEMs, i.e. \cite{vibe}
    \begin{equation}
        \frac{1}{L}\sum_{i=1}^L \frac{LLR^i}{\sqrt{\sum_{j=1}^{N_{sparse}}(LLR_j^i)^2}}.
    \end{equation}
\end{enumerate}

The core idea is that providing different propagated LLRs across the ensemble using different transfer matrices enriches the soft information used by the sparsified DEM to break distinct trapping sets. Since an ensemble is invoked, the diversity is then enriched by randomized serial BP schedules in a similar fashion as Vibe decoding. Furthermore, this randomization is further extended due to the two-stage BP nature of the decoder, where distinct schedules are employed for the full DEM and the sparsified one for each decoder across the ensemble. Thus, extra BP diversity for the Vibe-like ensembling is also achieved by the two-stage protocol.

As discussed, the DAWG ensemble is built by selecting a set of different transfer matrices that relate the full DEM with its sparsified version. Similar to the original proposal, we used a guided exhaustive search to find the transfer matrices \cite{otf}. 
The difference resides in the fact that we allow the search to look for higher order combinations of sparsified DEM columns to form the column of the DEM that is aimed to be decomposed. We then randomly chose one of the solutions for each of the transfer matrices, resulting in a set of transfer matrices with different weight combinations of sparsified columns forming ones of the DEM.

\section{Results}
In order to evaluate the performance of the DAWG+LSD-$0$ decoder, we performed numerical simulations (see Appendix \ref{app:num} for details) of BB and superdense color code circuits under standard depolarizing circuit-level noise. 

\subsection{Bivariate bicycle codes}
We began by considering the $[[72,12,6]]$, $[[108,8,10]]$ and $[[144,12,12]]$ BB codes \cite{BBcodes}. We employed the circuits implemented for these codes in \cite{bpgdg}. For these codes, we decode over the uncorrelated DEMs associated to the memory experiment. We employed the following BP setup to decode each of those:
\begin{itemize}
    \item An ensemble size of $10$ with $20$ total iterations for the $[[72,12,6]]$. DAWG uses $10$ iterations for the full DEM and $10$ for the sparsified DEM.
    \item An ensemble size of $15$ with $20$ total iterations for the $[[108,8,10]]$. DAWG uses $10$ iterations for the full DEM and $10$ for the sparsified DEM.
    \item An ensemble size of $20$ with $20$ total iterations for the $[[144,12,12]]$. DAWG uses $10$ iterations for the full DEM and $10$ for the sparsified DEM.
\end{itemize}

Notably, the sparsified models considered for the BB codes contain only $25\%$ of the variable nodes in the full DEMs (See Table \ref{tab:bbsizes}), and the mean degree of those is significantly lower, i.e. the sparsified DEMs are significantly sparser. We quantify the mean and maximum variable and check node degrees in Table \ref{tab:bbdegrees}. It is clear that the sparsified DEM is significantly less dense and with less node connectivity degree. This implies that both the second-stage BP and the LSD-$0$ post-processor must consider a DEM that is easier to decode over.

\begin{table}[h!]
\caption{Number of columns in the original and sparsified DEMs for BB codes. The DEM is the uncorrelated model. Recall that the number of rows is the same.}
\label{tab:bbsizes}
\begin{tabular}{lrr}
\hline
Code &  DEM columns  & SDEM columns  \\
\hline
$[[72,12,6]]$ & 2232 & 252  \\

$[[108,8,10]]$  & 5508 &  1728  \\

$[[144,12,12]]$  & 8784 &  2736  \\
\hline
\end{tabular}
\end{table}

\begin{table}[h!]
\caption{Check and variable node degrees for the full DEM and the sparsified DEM (SDEM) for BB codes. We show both the mean and maximum values. The SDEM shows a significantly sparser form.}
\label{tab:bbdegrees}
\begin{tabular}{lllll}
\hline
Node degree & DEM mean & DEM max  & SDEM mean & SDEM max \\
\hline
Check & 32.8 & 35 & 7.8 & 8 \\

Variable  & 3.5 &  6 & 2.7 & 3 \\
\hline
\end{tabular}
\end{table}

\begin{figure}
   a) $[[72,12,6]]$ BB code.\\ \includegraphics[width=0.9\linewidth]{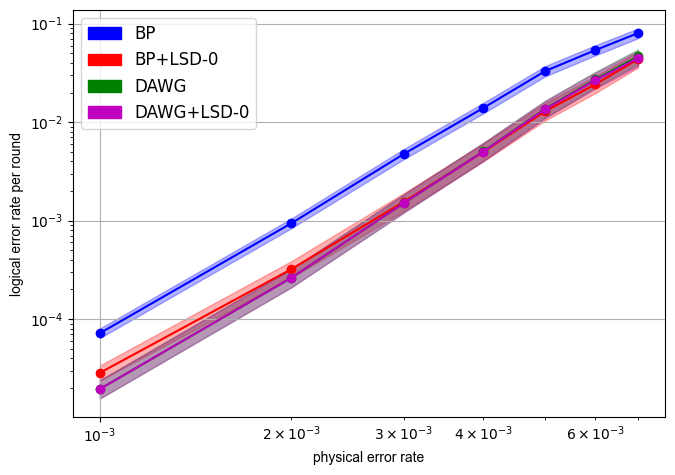}\\
    b) $[[108,8,10]]$ BB code.\\\includegraphics[width=0.9\linewidth]{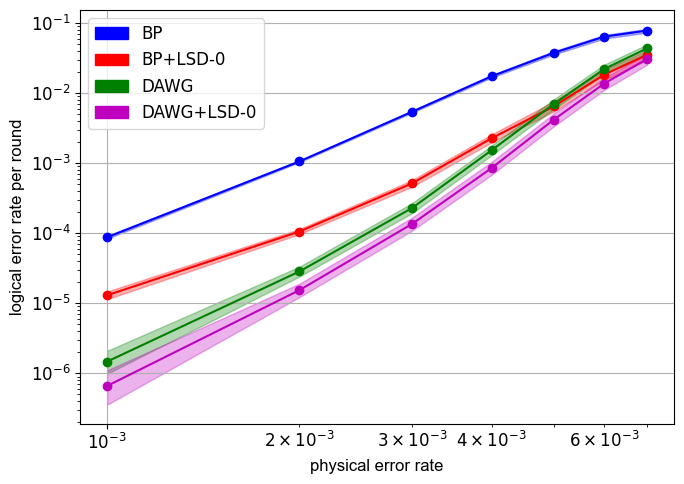}\\
     c) $[[144,12,12]]$ BB code.\\\includegraphics[width=0.9\linewidth]{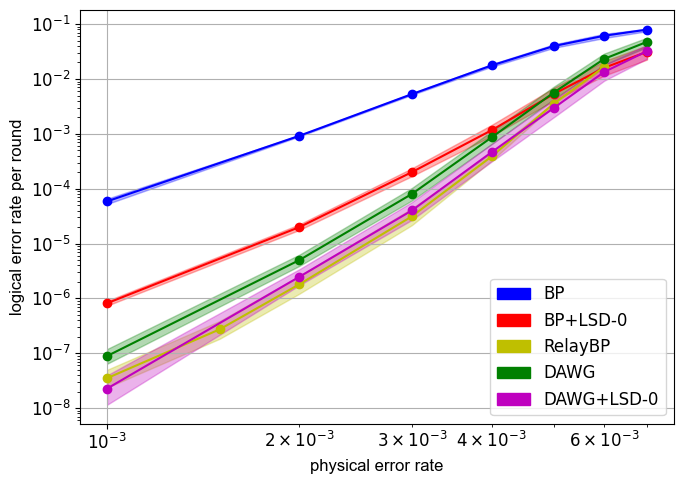}
    \caption{Performance of DAWG for BB codes. a) $[[72,12,6]]$. b) $[[108,8,10]]$. c) $[[144,12,12]]$. The shading highlights the region of estimated probabilities where the likelihood ratio is within a factor of $1000$.
    }
    \label{fig:BBresults}
\end{figure}

Figure \ref{fig:BBresults} showcases the obtained results compared to standalone serial BP, BP+LSD-$0$ \cite{joschkaLSD} and RelayBP \cite{relayBP}. The ldpc package \cite{joschkaLDPC} was used for the BP implementations. For RelayBP, we only showcase it for the Gross code and report the values obtained in \cite{gari} using the uncorrelated DEM. 

From the results, it can be seen that the performance of DAWG and DAWG+LSD-$0$ match those of BP+LSD-$0$ for the smallest BB code (Figure \ref{fig:BBresults}a). This comes from the fact that this code is non-degenerate and, thus, most of the errors are solved by means of BP. However, as the code distance increases, the BB codes become more degenerate and the true potential of DAWG can be observed. For the $[[108,8,10]]$ BB code shown in panel b) of Figure \ref{fig:BBresults}, it can already be seen that DAWG is able to operate an order of magnitude better than BP+LSD-$0$. Importantly, the LSD-$0$ post-processor does not seem to improve the logical error rate in a significant manner, indicating that DAWG can be run standalone for this code.

However, the main result here comes when considering the Gross code, which has been of central interest to the qLDPC community in recent times. The result depicted in panel c) of Figure \ref{fig:BBresults} showcases that DAWG is able to operate an order of magnitude better than BP+LSD-$0$ without the need for a post-processor. Furthermore, the use of an LSD-$0$ post-processor does show a significant boost for this code, reaching a logical error rate that matches that of RelayBP \cite{relayBP}. The latter algorithm has been considered as an important candidate for real-time decoding qLDPC codes and has been integrated in FPGA as a proof-of-concept \cite{relayfpga}. Thus, DAWG+LSD-$0$ is able to match the performance of the most accurate BP decoders in the literature. This is done without requiring an extensive optimization procedure such as the one required by the memory strengths that define the performance of RelayBP. This points towards the flexibility shown by DAWG to be able to decode, in principle, other noise models or even DEMs associated to logical operations successfully without needing extensive optimization \cite{galoisqudit}.

Nevertheless, it is important to comment that for the most degenerate codes, DAWG does seem to require some post-processing to reach the performance of the best decoders and to guarantee convergence to the observed syndromes. However, as it was discussed in \cite{fpgasbeverland}, if the probability to exceed the QEC cycle time is rare enough, the backlog that is generated by a potential latency tail can be effectively avoided for variable time decoding algorithms. In principle an integrated version of BP should be able to operate below QEC extraction cycles, so we will assume that the latency tail is generated when LSD-$0$ is invoked. Note how, for the Gross code, LSD-$0$ runs once per $10$ million decoding rounds at $p=0.1\%$ when considering the DAWG ensemble, i.e. it is used on very rare occasions.

\begin{figure}
\includegraphics[width=0.9\linewidth]{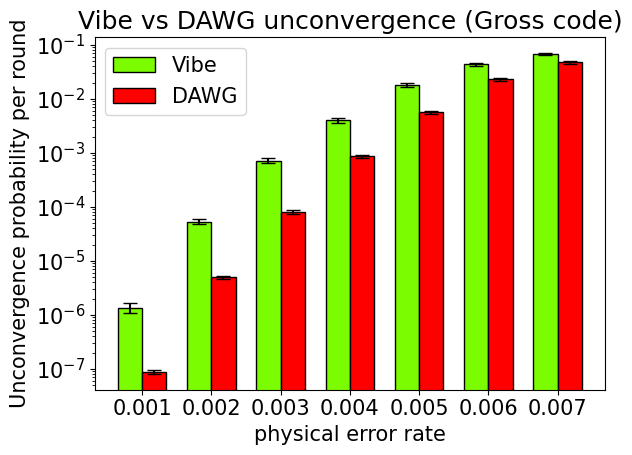}
    \caption{Probability that DAWG and Vibe do not converge per round for the Gross code. DAWG shows much rarer unconvergence than Vibe, with the difference being accentuated when the physical error rate lowers.
    }
    \label{fig:BBconvergence}
\end{figure}

To finish with BB codes, we showcase the convergence probability of DAWG when compared to Vibe decoding \cite{vibe}. Since DAWG does include BP ensembling and diversity ideas from Vibe, this comparison is integral to understand the benefit introduced by the ensemble of transfer matrices. Figure \ref{fig:BBconvergence} showcases the obtained results of the amount of shots per decoding round that do not converge for Vibe and DAWG considering the Gross code. Both use the same ensemble size as discussed before and the same total amount of serial BP iterations, i.e. $20$. The results showcase how DAWG consistently converges much more than Vibe, reaching to up to an order of magnitude less shots per round that are not solved for physical error rates below $p=0.004$. Furthermore, the gain introduced by DAWG becomes more noticeable when approaching practical QEC physical error rates around $p=0.001$. These results clearly indicate that the extra diversity introduced by DAWG makes the decoding proposal less reliant on a post-processor to reach good decoding performance and, thus, to show significantly lower latency tails. This is of paramount importance for real-time decoding.
\begin{table}[h!]
\caption{Number of columns in the original and sparsified DEMs for color codes. The DEM is the correlated model. Recall that the number of rows is the same.}
\label{tab:colorsizes}
\begin{tabular}{lrr}
\hline
Code &  DEM columns  & SDEM columns  \\
\hline
$d=3$ & 385 & 86  \\

$d=5$  & 2957 &  402  \\

$d=7$  & 9543 &  1083  \\

$d=9$  & 21995 &  2273  \\
\hline
\end{tabular}
\end{table}

\subsection{Superdense color code circuits}
We also applied the DAWG decoder to superdense color code circuits under standard circuit-level depolarizing noise to showcase the potential of the DAWG decoder \cite{chromobius}. 
Color codes are very interesting candidates for fault-tolerant quantum logic as they allow for the whole Clifford group in a transversal manner \cite{colorBombin}. Superdense circuits were proposed as an efficient way to do the syndrome extraction of these codes, but decoding them has shown to be challenging (similar to other color code implementations). Importantly, Vibe decoding was the first decoder to show logical error rates that make color codes competitive against surface codes \cite{vibe}. Artificial intelligence (AI) based predecoders have recently been used to complement the Chromobius decoder for color codes to approach the performance of Vibe \cite{chromobius,colorNvidia}.

We consider decoding color codes with distances $d=\{3,5,7,9\}$ implemented with superdense color code circuits. Unlike for BB codes, we decode over the correlated DEMs associated to memory experiments of these codes. We employ the following BP setup for each of these:

\begin{itemize}
    \item An ensemble size of $5$ with $20$ total iterations for the $d=3,5$. DAWG uses $10$ iterations for the full DEM and $10$ for the sparsified DEM.
    \item An ensemble size of $10$ with $20$ total iterations for the $d=7$. DAWG uses $10$ iterations for the full DEM and $10$ for the sparsified DEM.
    \item An ensemble size of $15$ with $20$ total iterations for the $d=9$. DAWG uses $10$ iterations for the full DEM and $10$ for the sparsified DEM.
\end{itemize}

As discussed, we decode this over the correlated DEMs for superdense color codes, whose sizes are significantly bigger and denser than the ones we used for BB codes. Crucially, the sparsified DEMs for these circuits contain only $10\%$ of the variable nodes of the full correlated DEMs (See Table \ref{tab:colorsizes}) while being significantly less dense. We quantify the variable and check node degrees in Table \ref{tab:colordegrees}, showing a more pronounced sparsification than for BB code circuits. This indicates that the sparsification routine can provide much more pronounced benefits for QEC circuits that involve more complicated DEMs. Note also that the LSD-$0$ post-processor will run over a much more restricted DEM than when using Vibe, providing a potential speedup for it as well.

\begin{table}[h!]
\caption{Check and variable node degrees for the full DEM and the sparsified DEM (SDEM) for superdense color code circuits. We show both the mean and maximum values. The SDEM shows a significantly sparser form.}
\label{tab:colordegrees}
\begin{tabular}{lllll}
\hline
Node degree & DEM mean & DEM max  & SDEM mean & SDEM max \\
\hline
Check & 189.2 & 270 & 10.1 & 14 \\

Variable  & 4.65 &  7 & 2.4 & 3 \\
\hline
\end{tabular}
\end{table}

\begin{figure}[h!]
\includegraphics[width=0.9\linewidth]{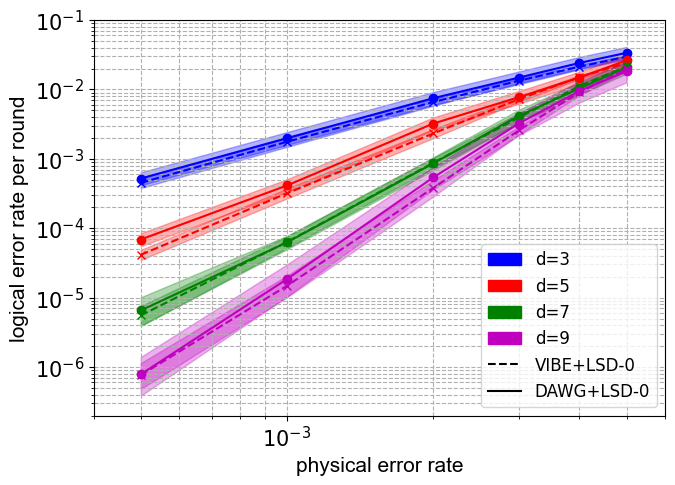}
    \caption{Performance of DAWG+LSD-$0$ for superdense color code circuits. The shading highlights the region of estimated probabilities where the likelihood ratio is within a factor of $1000$.
    }
    \label{fig:Colorresults}
\end{figure}

Figure \ref{fig:Colorresults} presents the obtained results for these superdense color code circuits using DAWG+LSD-$0$. A visual inspection shows that DAWG+LSD-$0$ is able to reach the same performance as Vibe+LSD-$0$ for the scenario considered. The data for Vibe+LSD-$0$ were obtained from the database associated with the original article \cite{vibe,vibezenodo}. Note that Vibe used a size $32$ ensemble to decode these circuits, while here the DAWG ensemble progressively uses significantly less, i.e. up to $15$. This indicates that DAWG+LSD-$0$ can potentially be an alternative to Vibe for color code decoding that requires a significantly smaller ensemble, critical to have decoders that are compact enough for specialized hardware integration. We do acknowledge that the work in \cite{vibe} probably did not aim to optimize for the smaller decoding ensemble, similar to our case here. However, the fact that the post-processor is run over the sparsified model when using DAWG significantly reduces the required latency and hardware footprint needed to run it, as we showcase later in the paper.

\begin{figure}
\includegraphics[width=0.9\linewidth]{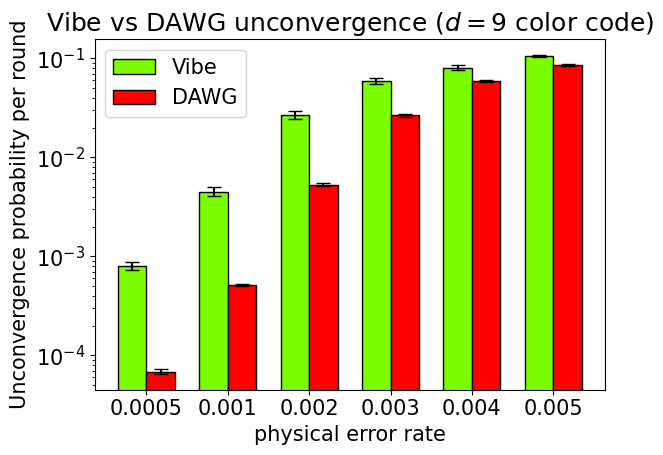}
    \caption{Probability that DAWG and Vibe do not converge per round for superdense color code circuit of $d=9$. DAWG shows much rarer unconvergence than Vibe, with the difference being accentuated when the physical error rate lowers.
    }
    \label{fig:Colorconvergence}
\end{figure}

In this line, and similar to our analysis for BB codes, we do also compare the unconvergence probability of DAWG when compared to Vibe under similar conditions. Figure \ref{fig:Colorconvergence} shows the probability per round that DAWG or Vibe do not converge considering the superdense color code circuit with distance $d=9$. Here, both Vibe and DAWG employ a size $15$ ensemble following the setup discussed previously, with $20$ total serial BP iterations. The results here are consistent with the ones obtained for the BB code and showcase that DAWG presents a fraction of the probability of Vibe to not converging below $p=0.002$. Note the gap between both decoders is accentuated when approaching physical error rates considered practical for QEC. This analysis points out that the DAWG ensemble makes invoking the LSD-$0$ post-processor an order of magnitude less times than Vibe, which is important to keep the latency tails introduced by it controlled. Furthermore, this analysis accentuates our previous discussion on the ensemble size, as the higher probability of convergence dictates that an smaller ensemble will be able to provide good decoding performance for color codes.

\subsection{The impact of DAWG in real-time implementations}

\begin{table*}[htbp]
  \centering
  \caption{Latency (per extraction round) in nanoseconds for different implementations. The jitter is on the scale of 100ps.}
  \label{tab:decoders}
  \begin{tabular}{lrrrrrr}
    \toprule
    Code & Rounds & Best case & Worst case DAWG & DAWG+OSD-$0$ & BP+OSD-$0$ (full DEM) \\
    \midrule
    $[\![72,12,6]\!]$   & 6  & 138.6   & 2,772   & 3,756 & 23,652\\
    $[\![108,8,10]\!]$  & 10  & 196.02 & 3,920.4
 & 5,324.4
 & 55,112.40 \\
    $[\![144,12,12]\!]$ & 12 & 257.4   & 5,148   & 6,996 & 86,652\\
    \midrule
    Color $d=3$ & 3 & 19.8  & 396  & 981.33  & 3,692\\
    Color $d=5$ & 5  & 59.4  & 1,188 & 1,725.6 & 25,924\\
    Color $d=7$ & 7 & 118.8 & 2,376 & 3,426.86 & 81,744\\
    Color $d=9$ & 9 & 198 & 3,960 & 5,690.22 & 186,400\\
    \bottomrule
  \end{tabular}
\end{table*}

In addition to the accuracy gains shown earlier in this section, DAWG is a hardware-friendly solution, as it combines two key benefits: (i) the transition between the full DEM and the sparsified one can be integrated into the control unit of the decoder, avoiding the implementation of two different cores or architectures; and (ii) the post-processing is executed on the sparsified DEM instead of the full one. The first benefit allows the serial BP architecture proposed in \cite{bascones2026scalablefpgaarchitecturerealtime} to be reused for both BP decoders, with the transfer matrices implemented through multiplexers at the inputs of the check-node units (CNUs), reducing the cost of implementing an ensemble. The second reduces the latency tails and the worst-case latency of the global decoding step, since the worst case of LSD-$0$ and OSD-$0$ post-processors is bounded by the number of error nodes, which is much larger in the full DEM than in the sparsified one.

As shown in Fig.~2 of \cite{otf}, the transition from the full DEM to the sparsified one keeps the number of check nodes constant and performs the reduction on the variable nodes, at the cost of combining the messages of several error nodes into one. To implement this procedure, the CNU inputs of a serial BP decoder, such as the one described in Fig.~2 of \cite{bascones2026scalablefpgaarchitecturerealtime}, require some extra logic that maps on the transfer matrix. During the first cycle of the sparsified BP, this logic selects, for each CNU, which inputs are combined to generate the priors of the sparsified graph. The combination sequence is stored in the control unit: each CNU selects among $d_c$ possible combinations, $d_c$ being the degree of the CNU, and updates the new priors of the corresponding error (or variable) nodes. At the same time, the unused messages are set to the maximum message value that can be represented, so that they do not contribute to subsequent computations, which indirectly masks the variable nodes that disappear in the transition between the full and the sparsified DEM. After this first iteration, the control is exactly the same as for the full graph, with the only exception that the vanished error/variable nodes are kept at the maximum message value, in other words, they are not updated. As a result, both decoders are integrated into a single architecture with a small overhead in the control.

For the uncorrelated graph, this architecture is very compact. In \cite{bascones2026scalablefpgaarchitecturerealtime}, 6 cores were implemented on a VCU19P FPGA \cite{amd_vu19p_fpga} together with the correlated part for the [[144,12,12]] code. It is therefore reasonable to assume that 12 cores can be implemented on a single chip at a maximum frequency of 300\,MHz if the correlated part is not considered. Since smaller BB codes have smaller DEM graphs, the same number of cores and the same maximum frequency can be expected for them. Consequently, the BP part of the DAWG decoder fits on a single FPGA for both [[72,12,6]] and [[108,8,10]] (the latter with minor optimizations), whereas two FPGAs would be required for [[144,12,12]].

Table~\ref{tab:decoders} reports the best-case latency, corresponding to a single BP iteration on the full DEM, and the worst-case latency, in which both BP decoders reach the maximum number of iterations on the full and on the sparsified DEM. We isolate the BP part of DAWG from the complete decoder including post-processing (full-DAWG) to show the impact of the second benefit, i.e., running LSD-$0$ or OSD-$0$ on the sparsified graph. For most of the reported codes, the worst case exceeds the 1\,$\mu$s latency constraint imposed by superconducting technology \cite{googleai_suppressing_2023} by at most a factor of 5.15. However, it is important to highlight that no architectural optimizations have been performed and the number of iterations could be adjusted without sacrificing too much accuracy. 

For the post-processing, we adopt OSD-$0$, since it achieves the same performance as LSD-$0$ while its architecture provides a worst-case latency that is deterministic, given by the number of error and check nodes. In other words, whenever the DAWG post-processor is invoked, the latency increase is deterministic for each code. We assume the architecture of \cite{bascones2025exploring} with its reported maximum frequency of 250\,MHz, obtained for the full DEM of the [[144,12,12]] code, and use the same frequency for the smaller codes. Since resources grow quadratically with the number of nodes, an optimized implementation on the sparsified DEM is expected to reach a much higher frequency, so the latencies in Table~\ref{tab:decoders} are upper bounds. Compared with BP+OSD-$0$ on the full DEM, DAWG reduces the worst-case latency by a factor between 7 and 40. This gap could grow even further, since the reported BP+OSD-$0$ implementation was optimized whereas DAWG was not. Thanks to the DAWG approach, the post-processor can work on the sparsified DEM alone, which is enough to achieve this large improvement. As a consequence, the latency distribution has shorter tails and the system behaves more consistently. Moreover, thanks to the compression of the DEM graph, adding the post-processor does not even double the worst-case latency of the BP processing with DAWG.

An interesting open question is the maximum code distance that the post-processor can support with this approach: since resources grow quadratically, distance 12 appeared to be the limit for current FPGAs with the full DEM \cite{bascones2025exploring}, \cite{maurya2026fpgatailoredalgorithmsrealtimedecoding}, whereas larger distances are expected to fit with the compressed/sparsified graph. Other promising directions include reducing latency with filtered OSD-$0$ at the cost of some accuracy \cite{maurya2026fpgatailoredalgorithmsrealtimedecoding}, or exploring efficient LSD-$0$ decoders with better average latency, even though their worst case would match that of OSD-$0$. Using other post-processors over the sparsified model informed by the BP runs on the full DEM is another interesting potential avenue to make those faster and more compact.

\section{Conclusions}
In this work, we have introduced decoding with alternating weighted graph sparsification (DAWG), a BP ensemble-based decoder for qLDPC codes. The decoder ensemble is built by finding different transfer matrices that relate a full DEM and a sparsified version of it \cite{otf}. DAWG is built to directly operate at the circuit-level by using a two-stage BP procedure based on DEM sparsification. Furthermore, it incorporates Vibe decoding ideas by randomizing all the schedules of the serial BP decoders across the ensemble \cite{vibe}. We numerically studied the performance of DAWG for BB and superdense color code circuits, observing excellent logical error rate performance. Specifically, we determined that DAWG is able to operate with the same performance as RelayBP when combined with an LSD-$0$ post-processor over the uncorrelated DEM of the Gross code. For superdense color codes considering the correlated DEM, we found that DAWG+LSD-$0$ is able to reach the performance of Vibe decoding, the leading proposal for color codes at the time of writing. All this is achieved with an order of magnitude better convergence rate than Vibe decoding under the same ensembling and BP iteration conditions. This has important consequences for real-time decoding since it simplifies the integration of the ensemble in specialized hardware and significantly reduces the impact of the latency tails introduced by the post-processor as well as reducing its baseline latency.

The key part of the DAWG decoder is the use of a BP+BP algorithm that uses mapped versions of the posterior LLRs of the BP run over the full DEM when it is not able to converge. The mapping is done by means of a transfer matrix that relates the variable nodes of both decoding graphs. In this work, we used a restricted exhaustive search, randomizing between solutions to find the transfer matrices integrated in the DAWG ensembles. The efficiency of this transfer matrix search technique for fault-tolerant quantum computing will strongly depend on the way in which the logic in such machine is conceived. Certain architectures conceive pre-computed DEMs for certain operations to then stich those in real-time, e.g. the LEGO architecture \cite{lego}, while others consider that the DEMs must be generated in real time. Thus, efficient ways of looking for transfer matrices are an important future line towards the application of DAWG or sparsification based decoder for full fault-tolerant logic. However, recall that matching decoders require this type of mapping from the full DEM to the matching graph, so it is highly likely that the ideas used in the surface code developments are adaptable to the qLDPC setting.

In this context, determining which sparsified models are best for these tasks is also of paramount importance. Recall that here we selected a quasi-phenomenological model as the sparsified model. This selection was originally motivated by the fact that many decoders excel over phenomenological models and fail to work in circuit noise. By informing the sparsified DEM with the posterior LLRs, many decoders can in principle operate correctly using two-stage procedures. However, the selection of a sparsified model is in principle arbitrary if its column space is the same as the one of the full DEM \cite{otf}. Therefore, proposing other sparsified DEM selections may lead to a better decoding performance or to a easier computation of the transfer matrix.

Another important aspect for the DAWG decoder is to showcase its efficiency in decoding logical operations. As discussed before, the DAWG decoder does not require complex parameter optimization to show excellent decoding performance, so it should translate the efficiency seamlessly to fault-tolerant logic. Furthermore, it is known that the decoding graphs involving generalized lattice surgery over qLDPC codes are usually denser and loopier than for memory experiments. Since sparsification was built to make such scenarios easier, DAWG and sparsification-based BP decoders could be very beneficial for such a setting. This should also translate to qLDPC codes with big and dense DEMs such as directional tile codes compiled in 2D lattices \cite{directionaltiles}. Recall the extreme size reduction and node degree reduction shown by the sparsified graphs in this work. This is critical for hardware implementation as, for example, the significant reduction in the number of columns is very beneficial for OSD-$0$ and LSD-$0$ as we discussed in this article.

\section{Acknowledgments}
We thank Antonio de Marti i Olius, Stergios Koutsioumpas and Joschka Roffe for many useful discussions and recommendations. We also thank Pedro Crespo for his guidance and the other members of the Quantum Information Group at Tecnun for their support.

This work was supported by the Spanish Ministry of Economy and Competitiveness through the MADDIE (Grant No. PID2022-137099NBC44) and GEMMA projects ``Generative AI and quantum approaches for advanced sensing and communications'' (Grant No. PID2025-169068NB-C33), and by the Ministry of Economic Affairs and Digital Transformation of the Spanish Government through the QUANTUM ENIA project call - QUANTUM SPAIN project, and by the European Union through the Recovery, Transformation and Resilience Plan - NextGenerationEU within the framework of the Digital Spain 2026 Agenda. We also acknowledge support from the Basque Quantum (BasQ) strategy through the ``Decoders for quantum low-density parity-check codes'' (qLDPCDec) project. F.G.H. acknowledges support from the project PID2023-147059OB-I00 funded by MCIU/AEI/ 10.13039/501100011033/ FEDER. UE.

\bibliographystyle{IEEEtran}
\bibliography{bibliography}

\begin{appendix}

\section{Numerical simulations}\label{app:num}
Monte Carlo computer simulations of the bivariate bicycle and superdense color code circuits have been performed with the objective of obtaining the performance curves (logical error rate). The circuit-level noise simulations have been done the following way. The sampling of the errors arising due to the noisy stabiliser circuit noise has been done by means of Stim \cite{stim}. We also use the SlidingWindowDecoding \cite{bpgdg,gitGuided} and the color-code-stim \cite{Lee2025colorcodedecoder,gitColor} Github repositories for constructing the Stim syndrome extraction circuits and obtain the circuit-level noise parity check and observable matrices. The decoder uses those to resolve the syndrome and return an error, which is later compared to the logical action of the Stim error.

The operational figure of merit we use to evaluate the performance of these quantum error correction schemes is the Logical Error Rate per syndrome cycle, i.e. the probability that a logical error has occurred after the recovery operation per syndrome extraction round \cite{BBcodes}.

Regarding the software implementations of the decoders used to perform the numerical simulations have been: the BP+LSD-$0$ implementation can be found in the LDPC Github package (v 2.1.9) \cite{Joschka_decoding_across,joschkaLDPC} while we implemented Vibe and DAWG using the BP decoders in such package. For the LSD-$0$ post-processor, the `bits\_per\_step' is set to $1$ for a progressive growth of the clusters.

As stated in the text, we employed randomized schedules for the serial BP decoders across the DAWG and Vibe ensembles. In this regard, our implementations randomize the schedules in each decoding shot, i.e. we do not look for deterministic serial schedules. The point of doing this is to show that the decoders work well on average. In practice, one should look for a set of schedules that work well in a deterministic manner. Note that LDPC v2 implements a vertical schedule (over variable nodes), while the horizontal schedule (over check nodes) is known to be faster as there are less detectors. Therefore, we consider that it is best to leave a deterministic implementation of DAWG and Vibe for practical scenarios and we refer to the average performance of them in this algorithmic paper.

Furthermore,  we set the clipping range at $\mathrm{LLR}\in[-80,80]$ for the LLR probability transfer from full DEMs to saprsified ones for numerical stability.

\end{appendix}
\end{document}